\documentclass[11pt]{revtex4}    
\usepackage{amssymb,epsf}
\usepackage{latexsym}
\usepackage{hyperref,natbib,textcase,bm}
\usepackage{amsmath,mathrsfs}
\usepackage{graphicx,epsfig}
\usepackage{latexsym,amssymb}
\usepackage{epsf,longtable,array}
\begin{document}

\title{Slowly Rotating Black Holes in Brans-Dicke-Maxwell Theory}
\author{H. Alavirad  \footnote{alavirad@particle.uni-karlsruhe.de}}
\affiliation{Institute for Theoretical Physics, University of Karlsruhe, Karlsruhe Institute of Technology, 76128 Karlsruhe, Germany}
\begin{abstract}
In this paper, we construct a class of (n+1)-dimensional $(n\geq4)$
slowly rotating black hole solutions in Brans-Dicke-Maxwell theory with a quadratic
potential.
These solutions can represent black holes with inner and outer event horizons, an extreme
black hole and a naked singularity and they are neither asymptotically
flat nor (anti)-de Sitter. We compute the Euclidean action and use it to obtain
the conserved and thermodynamics quantities such as entropy, which does not obey the area law.
We also compute the angular momentum and the gyromagnetic ratio for these type
of black holes where the gyromagnetic ratio is modified in Brans-Dicke theory
 compared to the Einstein theory.

\end{abstract}

\maketitle
\section{introduction\label{I}}

 Brans-Dicke(BD) theory \cite{Jordan} is perhaps the most common alternative theory to the
Einstein's general relativity. This theory contains both Mach's principle and Dirac's
large number hypothesis. The theory has recently received interest as it arises naturally as
the low energy limit of many theories of quantum gravity such as the supersymmetric string theory or
 Kaluza-Klein theory, and is also found to be consistent with present cosmological observations \cite{La}. The
theory contains an adjustable parameter ,$\omega$, that represents the strength of the coupling between matter
 and the scalar field. For certain values of $\omega$, the BD theory agrees with GR in the post-Newtonian limit up
to any desired accuracy and hence weak-field observations cannot rule out the BD theory in favor of
general relativity \cite{Will}, although the singularity problem remains.

Shortly after the appearance of this theory one of its authors, C. Brans, obtained the statically spherically
symmetric solutions \cite{Brans}. Since then many authors
have investigated black holes in Brans-Dicke theory \cite{Sen}. Hawking
proved in four dimensions the stationary and vacuum Brans-Dicke solution is just the Kerr solution
with a constant scalar field everywhere \cite{Hawking}. Cai and Myung have proved that
in four dimensions, the charged black hole solution in the Brans-Dicke-Maxwell theory
is just the Reissner-Nordstrom solution with a constant scalar field \cite{Cai}. The
Kerr-Newman type black hole solutions, which are different from the solutions in general relativity, 
 have been constructed for $-5/2<\omega<-3/2$ in \cite{Kim}. Thermodynamics
of black holes in Brans-Dicke theory is investigated \cite{Kim2}.

\par On the other hand, the rotating black hole solutions in higher dimensional
Einstein gravity was found by Myers and Perry \cite{Myers3}. The solutions were
uncharged and can be considered as a generalization of the four dimensional Kerr solutions.
Moreover it has recently been shown that the gravity in higher dimensions contains much richer
dynamics than in four dimensions. As an example, there exists a black ring solution in five dimensions
with the horizon topology of $S^2\times S^2$ \cite{Emparan}, which has
the same mass and angular momentum as the Myers-Perry solution and therefore
contradicts the uniqueness theorem in five dimensions. Although the non-rotating
black hole solutions in higher dimensional Einstein-Maxwell gravity was found
 many years ago \cite{Tangh}, the analytic solution of a generalization of
 the charged Myers-Perry solution in $(n+1)$-dimensional Einstein Maxwell gravity has not been found yet.
 Solutions of different kinds of charged rotating black holes in
 higher dimensions have been discussed in the framework of
 supergravity and string theory \cite{Guan, Youm, Chong}. In \cite{Ali}, the solutions of
 charged rotating black hole in $(n+1)$-dimensional Einstein-Maxwell
 theory with a single rotation parameter in the limit of slow rotation
 have been constructed. In addition, \cite{Hye} contains a class of charged
 slowly rotating black hole solutions in Gauss-Bonnet gravity. Rotating black
  holes in Einstein-Maxwell-Dilaton gravity is discussed in \cite{sheykhi2}.

\par In this paper we investigate charged slowly rotating black holes in Brans-Dicke theory by using
the conformal transformation between dilaton fields and Brans-Dicke theory. The structure of this paper is as follows:
In section \ref{II} we obtain the solution of charged rotating Brans-Dicke black holes and discuss
their causal structure, then
in section \ref{III} we obtain the conserved
quantities of the finite action by using the Euclidean action and section \ref{IV} contains our results.

\section{Slowly Rotating Black Holes in Brans-Dicke-Maxwell theory \label{II}}
The action of the Brans-Dicke-Maxwell theory in $(n+1)$-dimensions with a scalar field $\Phi$ and
a self-interacting potential $V(\Phi)$ is
\begin{equation}
I_{G}=-\frac{1}{16\pi}\int_{\mathcal{M}}d^{n+1}x\sqrt{-g}(\Phi \mathcal{R}-\frac{\omega}{\Phi}(\nabla\Phi)^2
-V(\Phi)-F_{\mu \nu}F^{\mu \nu})\label{1}
\end{equation}
where $\mathcal{R}$ is the Ricci scalar, $F_{\mu \nu}=\partial_{\mu}A_{\nu}-\partial_{\nu}A_{\mu}$
 the electromagnetic tensor field, $A_{\mu}$ the vector potential, $\omega$ 
the coupling constant and $\Phi$ is the BD scalar field. By varying the the action
(\ref{1}) with respect to the metric $g_{\mu \nu}$, the scalar field $\Phi$ and
vector field $A_{\mu}$, one can obtain the
following field equations
\begin{eqnarray}
G_{\mu \nu}&=&\frac{\omega}{\Phi^2}(\nabla_{\mu}\Phi\nabla_{\nu}\Phi-\frac{1}{2}g_{\mu \nu}(\nabla\Phi)^2)
-\frac{V(\Phi)}{2\Phi}g_{\mu \nu}+\frac{1}{\Phi}(\nabla_{\mu}\nabla_{\nu}-g_{\mu \nu}\nabla^2\Phi) \nonumber \\
&+&\frac{2}{\Phi}(F_{\mu \lambda}F_{\nu}^{\Lambda}-\frac{1}{2}F_{\rho \sigma}F^{\rho \sigma}g_{\mu \nu})
\label{2}
\end{eqnarray}
\begin{equation}
\nabla^2\Phi=-\frac{n-3}{2[(n-1)\omega+n]}F^2+\frac{1}{2[(n-1)\omega+n]}\left[(n-1)\Phi\frac{dV(\Phi)}{d\Phi}
-(n+1)V(\phi)\right]\label{3}
\end{equation}
\begin{equation}
\nabla_{\mu}F^{\mu \nu}=0\label{4}
\end{equation}
where $G_{\mu \nu}$ is the Einstein tensor. It is not easy to solve
the field equations (\ref{2})-(\ref{4}) directly because
the right hand side of
eq. (\ref{2}) includes the second derivatives of the scalar field. Fortunately,
we can transform these field equations to the dilaton field equations by
conformal transformations. If one uses the following conformal transformations
\begin{eqnarray}
\bar{g}_{\mu \nu}=\Phi^{2/(n-1)}g_{\mu \nu} \nonumber \\
\bar{\Phi}=\frac{n-3}{4\alpha}ln\Phi \label{5}
\end{eqnarray}
where
\begin{equation}
\alpha=(n-3)/\sqrt{4(n-1)\omega+4n} \label{6}
\end{equation}
then the action (\ref{1}) takes the form
\begin{equation}
\bar{I}_{G}=-\frac{1}{16\pi}\int_{\mathcal{M}}d^{n+1}x\sqrt{-\bar{g}}\{\mathcal{\bar{R}}-\frac{4}{n-1}(\bar{\nabla}\
\bar{\Phi})^2-\bar{V}(\bar{\Phi})-e^{-\frac{4\alpha\bar{\Phi}}{n-1}}\bar{F}_{\mu \nu}\bar{F}_{\mu \nu}\}\label{7}
\end{equation}
where $\mathcal{\bar{R}}$ and $\bar{\nabla}$ are the Ricci scalar and covariant derivative corresponding to the metric $\bar{g}_{\mu \nu}$
and $\bar{V}(\bar{\Phi})$ is:
\begin{equation}
\bar{V}(\bar{\Phi})=\Phi^{-(n+1)/(n-1)}V(\Phi)\label{8}
\end{equation}
\par Eq. (\ref{7}) is simply the action of (n+1)-dimensional Einstein-Maxwell-dilaton
 gravity, where $\bar{\Phi}$ is the dilaton field, $\bar{V}(\bar{\Phi})$ a potential
for $\bar{\Phi}$ and $\alpha$ is a constant that determines the strength
of coupling of the scalar and electromagnetic field $\bar{F}_{\mu \nu}$.
By varying  the action (\ref{7}) with respect to $\bar{g}_{\mu \nu}$
, $\bar{\Phi}$ and $\bar{F}_{\mu \nu}$, we obtain
\begin{equation}
\bar{\mathcal{R}}_{\mu \nu}=\frac{4}{n-1}(\bar{\nabla}_{\mu}\bar{\Phi}\bar{\nabla}
_{\nu}\bar{\Phi}+\frac{1}{4}\bar{V}(\bar{\Phi})\bar{g}_{\mu \nu})+
2e^{-4\alpha\bar{\Phi}/(n-1)}(\bar{F}_{\mu \lambda}\bar{F}_{\nu}^{\lambda}
-\frac{1}{2(n-1)}\bar{F}_{\rho \sigma}\bar{F}^{\rho \sigma}\bar{g}_{\mu \nu})
\label{9}
\end{equation}
\begin{equation}
\bar{\nabla}^2\bar{\Phi}=\frac{n-1}{8}\frac{\partial\bar{V}}{\partial\bar{\Phi}}
-\frac{\alpha}{2}e^{-4\alpha\bar{\Phi}/(n-1)}\bar{F}_{\rho \sigma}\bar{F}^{\rho \sigma}
\label{10}
\end{equation}
\begin{equation}
\bar{\nabla}_{\mu}[\sqrt{-\bar{g}}e^{-4\alpha\bar{\Phi}/(n-1)}\bar{F}^{\mu \nu}]=0
\label{11}
\end{equation}
 Many authors have obtained the solutions of above field equations \cite{sheykhi1,Gibb, Garf, Mann, Pol, Cai5, Cle, Yaz, sheykhi2}.
It is now simple to obtain the solutions of field equations (\ref{2})-(\ref{4}) by applying
the conformal transformations (\ref{5}) to the solution of the field equations (\ref{9})-(\ref{11}).
In \cite{sheykhi2}, the solution of field equations (\ref{9})-(\ref{11})
has been obtained in the slowly rotating case where $a\ll 1$. In this
 case the only term in the metric which changes to $O(a)$ is $g_{t\phi}$, and $A_\phi$ is the only component
 of the vector potential that changes, where the dilaton field does not change to $O(a)$. Therefore,
 for infinitesimal angular momentum up to $O(a)$, we can take the following form for the metric in
 (n+1)-dimension for Einstein-Maxwell-dilaton theory \cite{sheykhi2}

 \begin{eqnarray}
 ds^2=&-&U(r)dt^2+\frac{dr^2}{U(r)}-2af(r)sin^2\theta dtd\phi\\ \nonumber
 &+&r^2R^2(r)(d\theta^2+sin^2\theta d\phi^2+cos^2\theta d\Omega^2_{n-3}) \label{12}
 \end{eqnarray}
where U(r), R(r) and f(r)  are functions of $r$, and $a$ is a parameter
associated with its angular momentum and $d\Omega^2_{n-3}$ denotes the
metric of an unit $(n-3)$-sphere. For small values of $a$, U(r) is a function only of $r$. From equation (\ref{11}) we can obtain the $t$ component of
the Maxwell equations
\begin{equation}
\bar{F}_{t r}=\frac{qe^{4\alpha\bar{\Phi}/(n-1)}}{(rR)^{n-1}}
\label{13}
\end{equation}
where q is an integration constant related to the electric
charge of the black hole. By using the definition $Q=\frac{1}{4\pi}\int exp[-4\alpha\bar{\Phi}/(n-1)]Fd\Omega$
we obtain the electric charge as
\begin{equation}
Q=\frac{q\omega_{n-1}}{4\pi}
\label{14}
\end{equation}
where $\omega_{n-1}$ represents the volume of a hypersurface with constant curvature.
In general, when we have a rotational parameter, there is also a vector potential
of the form
\begin{equation}
A_{\phi}=aqh(r)sin^2\theta
\label{14.1}
\end{equation}

 In \cite{sheykhi2}, for a Liouville-type dilaton potential, is introduced  for $\bar{V}(\bar{\Phi})$
\begin{equation}
\bar{V}(\bar{\Phi})=2\Lambda^{2\zeta\bar{\Phi}}
\label{15}
\end{equation}
where $ \Lambda$ and $\zeta$ are constants. The
unknown functions $U(r),f(r)$ and $R(r)$ are obtained using the ansatz in \cite{sheykhi2}
\begin{equation}
R(r)=e^{2\alpha\bar{\Phi}/(n-1)}
\label{16}
\end{equation}
By substituting eq. (\ref{16}), the Maxwell fields (\ref{13}) and (\ref{14.1}) and the
metric (\ref{12}) into the field equations (\ref{9})-(\ref{11}), we can obtain
\begin{eqnarray}
U(r)&=&-\frac{(n-2)(\alpha^2+1)^2b^{-2\gamma}r^{2\gamma}}{(\alpha^2-1)(\alpha^2+n-2)}-
\frac{m}{r^{(n-1)(1-\gamma)-1}}+
\frac{2q^{2}(\alpha^2+1)^{2}b^{-2(n-2)\gamma}}{(n-1)(\alpha^2+n-2)}r^{2(n-2)(\gamma-1)} \nonumber \\
\label{17}
\end{eqnarray}
\begin{equation}
f(r)=\frac{m(\alpha^2+n-2)b^{(n-3)\gamma}}{\alpha^2+1}r^{(n-1)(n-\gamma)+1}
-\frac{2q^2(\alpha^2+1)b^{(1-n)\gamma}}{n-1}r^{2(n-2)(\gamma-1)}
\label{17.1}
\end{equation}
\begin{equation}
\bar{\Phi}=\frac{(n-1)\alpha}{2(1+\alpha^2)}ln(\frac{b}{r})
\label{18}
\end{equation}
\begin{equation}
h(r)=r^{(n-3)(\gamma-1)-1}
\label{18.1}
\end{equation}

where $\gamma=\alpha^2/(1+\alpha^2)$ and b is an arbitrary constant.
In addition we should have
\begin{equation}
\zeta=\frac{2}{\alpha(n-1)}, \hspace{1cm}
\Lambda=\frac{(n-1)(n-2)\alpha^2}{2b^2(\alpha^2-1)}
\label{20}
\end{equation}
in order to fully satisfy the field equations.
\par To obtain the solutions of the field equations (\ref{2})-(\ref{4}) in the Brans-Dicke-Maxwell theory,
we take a metric of the form
\begin{eqnarray}
 ds^2=&-&A(r)dt^2+\frac{dr^2}{B(r)}-2ag(r)sin^2\theta dtd\phi\\ \nonumber
 &+&r^2H^2(r)(d\theta^2+sin^2\theta d\phi^2+cos^2\theta d\Omega^2_{n-3}) \label{21}
\end{eqnarray}
To determine the unknown functions $A(r), B(r), g(r)$ and $H(r)$ we apply the
conformal transformations (\ref{5}), (\ref{6}) and (\ref{8}) to eqs. (\ref{16}),(\ref{17}) and (\ref{18}).
Leading to
\begin{eqnarray}
A(r)&=&-\frac{(n-2)(\alpha^2+1)^2b^{-2\gamma(\frac{n-5}{n-3})}r^{2\gamma(\frac{n-5}{n-3})}}
{(\alpha^2-1)(\alpha^2+n-2)}
 \nonumber \\ &+&
\frac{2q^{2}(\alpha^2+1)^{2}b^{2\gamma(2-n+\frac{2}{n-3})}}{(n-1)(\alpha^2+n-2)}r^{2(n-2)(\gamma-1)-
\frac{4\gamma}{n-3}}
-
\frac{mb^{\frac{4\gamma}{n-3}}r^{\gamma(n-1-\frac{4}{n-3})}}{r^{(n-2)}}
\label{22}
\end{eqnarray}
\begin{eqnarray}
B(r)&=&-\frac{(n-2)(\alpha^2+1)^2b^{-2\gamma(\frac{n-1}{n-3})}r^{2\gamma(\frac{n-1}{n-3})}}
{(\alpha^2-1)(\alpha^2+n-2)}
\nonumber \\ &+&
\frac{2q^{2}(\alpha^2+1)^{2}b^{-2\gamma(n-2+\frac{2}{n-3})}}{(n-1)(\alpha^2+n-2)}r^{2(n-2)(\gamma-1)+
\frac{4\gamma}{n-3}}
-\frac{mb^{-\frac{4\gamma}{n-3}}r^{\gamma(n-1+\frac{4}{n-3})}}{r^{(n-2)}}
\label{23}
\end{eqnarray}
and
\begin{equation}
g(r)=\frac{m(\alpha^2+n-2)b^{(n-3+\frac{4}{n-3})\gamma}}{\alpha^2+1}r^{(n-1)(n-\gamma)+1-\frac{4\gamma}{n-3}}
-\frac{2q^2(\alpha^2+1)b^{(1-n+\frac{4}{n-3})\gamma}}{n-1}r^{2[(n-2)(\gamma-1)-\frac{2\gamma}{n-3}]}
\label{23.1}
\end{equation}
\begin{equation}
H(r)=(\frac{b}{r})^{\gamma\frac{n-5}{n-3}}
\label{24}
\end{equation}
\begin{equation}
\Phi(r)=(\frac{b}{r})^{2\gamma\frac{n-1}{n-3}}
\label{24}
\end{equation}
\begin{equation}
F_{tr}=\frac{qb^{\gamma(3-n)}}{r^{(n-3)(1-\gamma)+2}}
\label{25}
\end{equation}

\begin{equation}
V(\Phi)=2\Lambda\Phi^{\frac{1}{\alpha(n-2)}[(\alpha+1)-4]}
\label{26}
\end{equation}
\begin{figure}[b]
\begin{minipage}{7.0 cm}
\includegraphics[width=7.0 cm]{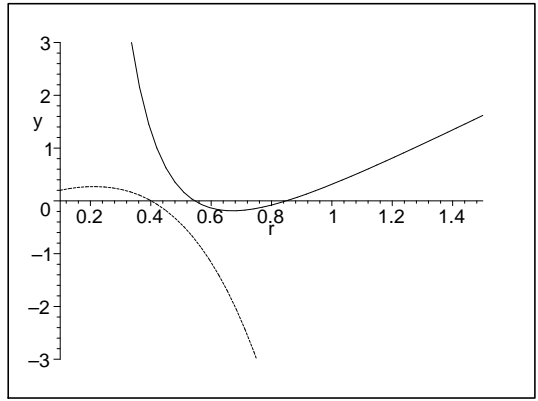}
\caption{The function B(r) versus r for $n=4$,
$m=2$, $b=1$ and $q=1$. $\alpha=0.5$ (bold line) and $\alpha=1.2$ (dashed line)}\label{1}
\end{minipage}
\end{figure}
\par At this point, it is worthwhile to investigate the
physical properties of these solutions. We can show that
the Kretschmann scalar $R_{\mu \nu \lambda \kappa}R^{\mu \nu \lambda \kappa}$
diverges at $r=0$, and it is finite for $r\neq0$ and approaches to zero as $r\rightarrow0$.
We find that there is an essential singularity at r=0.
In addition, the solution is ill-defined for $\alpha=1$ and the
cases $\alpha>1$ and $\alpha<1$ should be considered separately. For
 $\alpha>1$, there exists a cosmological horizon (fig. 1) whereas
there is no cosmological horizon for $\alpha<1$. In this case 
eq. (\ref{22}) contains a wide range of causal structure which depends on the values
of the metric parameters $\alpha, m, q$ and $k$ (fig. 2-3).
\par Moreover we can obtain some information about
causal structure by considering the temperature of the horizons.
 By using the definition of Hawking temperature on the outer
horizon $r_+$, which may be obtained through the definition of surface gravity
\begin{figure}[t]
\begin{minipage}{7.0 cm}
\includegraphics[width=7.0 cm]{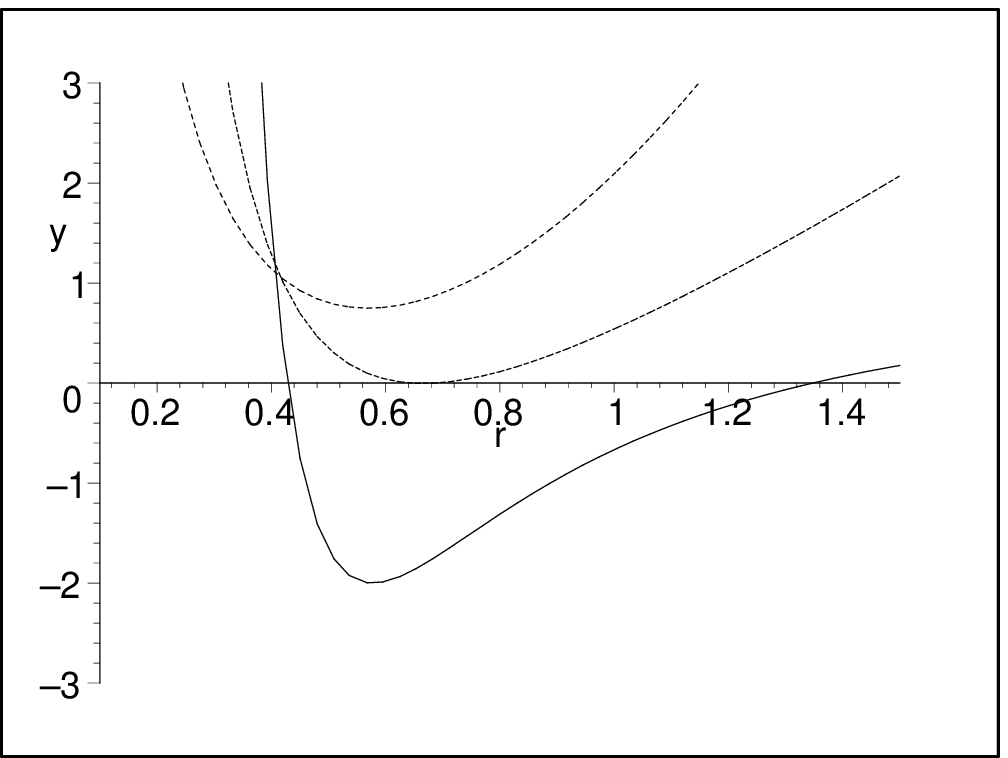}
\caption{The function B(r) versus r for $n=4$,
$m=2$, $b=1$ and $q=1$, $\alpha=0$ (bold line), $\alpha=0.54$ (dashed line)
and $\alpha=0.7$ (dotted line)}\label{2}
\end{minipage}
\hfill
\begin{minipage}{7.0 cm}
\includegraphics[width=7.0 cm]{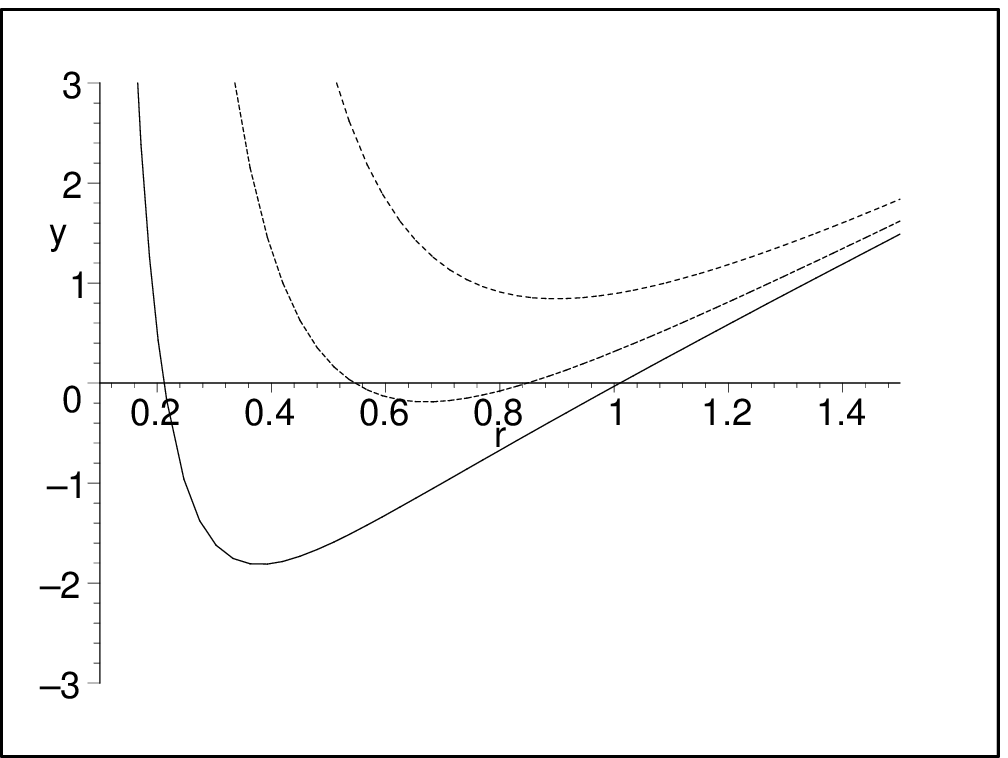}
\caption{The function V(r) versus r for $n=4$,
$m=2$ and $b=1$, $\alpha=0.5$ $q=0.5$(bold line), $q=1$ (dashed line)
and $q=1.5$ (dotted line)}\label{3}
\end{minipage}
\end{figure}
\begin{equation}
T_{+}=\frac{1}{2\pi}\sqrt{-\frac{1}{2}(\nabla_{\mu}\chi_{\nu})(\nabla^{\mu}\chi^{\nu})}
\label{27}
\end{equation}
where $\chi$ is the Killing vector $\partial_{t}$, we can write
\begin{eqnarray}
T_{+}&=&-\frac{(\alpha^2+n-2)m}{4\pi(\alpha^2+1)}r_{+}^{(n-1)(\gamma-1)}
+\frac{(n-2)(\alpha^2+1)b^{-2\gamma}}{2\pi(1-\alpha^2)}r_{+}^{2\gamma-1}
\label{28}
\end{eqnarray}
We see from the above equation that the temperature is invariant under
conformal transformations \cite{sheykhi2}, which is a result of the regularity of the
conformal parameter at the horizon.
For $\alpha>1$ we find that the temperature is negative from eq. (\ref{28}).
Numerical calculations show that the temperature of the event horizon
goes to zero as the black hole approaches an extreme one. In addition we can
show that for $(\alpha<1)$
\begin{equation}
m_{ext}=\frac{2(n-2)(\alpha^2+1)^2b^{-2\gamma}}{(n-\alpha^2)
(\alpha^2+n-2)}r_{+}^{(2-n)(\gamma-1)+\gamma}
\label{29}
\end{equation}
In summary, the metric (\ref{21}) can represent a rotating black hole with
two inner and outer horizons located at $r_{+}$ and $r_{-}$ provided that the mass
parameter m is greater than $m_{ext}$, an extreme black hole when $m=m_{ext}$, and a naked singularity
when $m<m_{ext}$.
\par The electric potential $U$, measured at infinity with respect to the horizon,
 is defined by \cite{Cve}
 \begin{equation}
 U=A_{\mu}\chi^{\mu}\mid_{\rightarrow\infty}-A_{\mu}\chi^{\mu}\mid_{\rightarrow 0}\label{U0}
 \end{equation}
where $\chi$ is the null generator of the event horizon. Therefore we obtain:
\begin{equation}
U=\frac{qb^{\gamma(3-n)}}{\Xi\Gamma r_{+}^{\Gamma}}\label{U}
\end{equation}
where $\Gamma=\gamma(3-n)+n-2$.

\section{Euclidean action and conserved quantities \label{III}}
The ADM (Arnowitt-Deser-Misner) mass $M$, entropy $S$ and electric
potential $U$ of the topological black hole
can be calculated through the use of the Euclidean action method
\cite{Cai4}.
In this approach, the electrical potential and
the temperature are initially  fixed on a boundary with a fixed radius
$r_{+}$. The Euclidean action has two parts; bulk and surface.
The first step to make the Euclidean action is to substitute $t$ with
$i\tau$ (this change does not affect our physical parameters, especially the angular
 momentum, see eq. (\ref{J})). This makes the metric positive definite:
\begin{eqnarray}
 ds^2=&+&A(r)dt^2+\frac{dr^2}{B(r)}-2ag(r)sin^2\theta dtd\phi\\ \nonumber
 &+&r^2H^2(r)(d\theta^2+sin^2\theta d\phi^2+cos^2\theta d\Omega^2_{n-3}) \label{posmet}
\end{eqnarray}
There is a conical singularity at the horizon $r=r_{+}$ in the Euclidean metric.
To eliminate it, the Euclidean time $\tau$ is made periodic with period $\beta$, where
$\beta$ is the inverse of Hawking temperature. We can now calculate the Euclidean action of
$(n+1)$-dimensional Brans-Dicke-Maxwell theory. It can be obtained analytically and by
continuously  changing of action (\ref{1}) to Euclidean time $\tau$, i.e.,
\begin{equation}
I_{GE}=-\frac{1}{16\pi}\int_{\mathcal{M}}
d^{n+1}x\sqrt{g}\left(\Phi {R}-\frac{\omega}{\Phi}(\nabla\Phi)^2
-V(\Phi)-F_{\mu \nu}F^{\mu \nu}\right)+
\frac{1}{8\pi}\int{d^{n}x\sqrt{h}\Phi(K-K_{0})},\label{act1e}
\end{equation}
where $K$ represents the extrinsic curvature on the induced metric $h$,
and $K_{0}$  is the extrinsic curvature on  the metric $h$ for flat space-time,
which must be added so that the Euclidean action is normalized to zero
in flat space-time. Using the metric (\ref{posmet}), we find to $O(1)$
\begin{eqnarray}
&&R=-g^{-1/2}(g^{1/2}U^{\prime}V/U)^{\prime}-2G^{0}_{0}+\textsc{O}(a^2)+\textsc{O}(a^4) \nonumber \\
&&K=-\frac{\sqrt{B(r)}}{2}\frac{A^{\prime}(r)rH(r)+2(n-2)A(r)H(r)+(2n-7)A(r)rH^{\prime}(r)}{A(r)rH(r)} \nonumber\\
&&K_{0}=\frac{2(n-3)-\gamma}{2r}\sqrt{\frac{(n-2)(\alpha^2+1)^2}{(\alpha^2-1)(\alpha^2+n-2)}}(\frac{r}{b})^{-\frac{n-1}{n-3}}\label{RE}
\end{eqnarray}
where $G^{0}_{0}$ is the 0-0 component of the Einstein tensor. By
substituting eq. (\ref{RE}) in action (\ref{act1e}) and using
eqs. (\ref{22})-(\ref{26}), we obtain
\begin{eqnarray}
I_{GE}&=&\beta\frac{\omega_{n-1}}{16\pi}\left( \frac{b^{(n-1)\gamma}(n-1)}{(\alpha^2+1)}\right)
-\frac{\omega_{(n-1)}}{4l^{n-2}}\left(b^{(n-1)\gamma}r_{+}^{(n-1)(1-\gamma)}\right)\nonumber \\
&-&\beta\frac{\omega_{n-1}}{8\pi }\frac{(n-\alpha^2)(\alpha^2+n-2)b^{2(n-2)\gamma}\omega_{n-1}}
{n(n-2)(\alpha^2+1)^2}ma^2-\beta\frac{q^2}{8\pi\Gamma r_{+}^{\Gamma}}
\label{IE}
\end{eqnarray}
where $\Gamma=(n-3)(1-\gamma)+1$. According to Ref. \cite{BRO}, the
thermodynamical potential can be given by $I_{GE}$
\begin{equation}
I_{GE}=\beta M-S-\beta Uq-\beta\Omega J
\label{GD}
\end{equation}
where $M$ is the ADM mass, $S$ the entropy and, $U$ the electric potential
corresponding to the conservation of charge q and $\Omega=a$ in this case. Comparing eqs. (\ref{IE}) and (\ref{GD}), we find
\begin{equation}
{M}=\frac{b^{(n-1)\gamma}}{16 \pi}\frac{n-1}{1+\alpha^2}\omega_{(n-1)}m,
\label{Mass}
\end{equation}
\begin{equation}
S=\frac{b^{(n-1)\gamma}r_{+}^{(n-1)(1-\gamma)}}{4l^{(n-2)}}\label{entropy}
\end{equation}
and
\begin{equation}
Q=\frac{q\omega_{(n-1)}}{4\pi}\label{charge}
\end{equation}
\begin{equation}
J=\frac{(n-\alpha^2)(\alpha^2+n-2)b^{2(n-2)\gamma}\omega_{n-1}}
{8\pi n(n-2)(\alpha^2+1)^2}ma \label{J}
\end{equation}

We can see from the above equations that the ADM mass, entropy and electric potential
are invariant under the conformal transformation
\cite{sheykhi1}. In addition, in the context of BD gravity, where we have the additional
gravitational scalar degree of freedom, the entropy of the black
hole does not follow the area law. This is due to the fact that
the black hole entropy comes from the boundary term in the
Euclidean action formalism.

 In addition,
the charge which is calculated in eq. (\ref{charge}) is the same as the one which was calculated in eq. (\ref{14}). By combining
eqs. (\ref{Mass}) and (\ref{J}), we can write:
\begin{equation}
J=\frac{2Ma}{n-1}\label{JJ}
\end{equation}
In order to calculate the gyromagnetic ratio of this type of black hole, we first need the magnetic dipole
moment for slowly rotating black holes, i.e., $\mu=Qa$, then the gyromagnetic ratio is given
by
\begin{equation}
g=\frac{2\mu M}{QJ}=\frac{n(n-1)(n-2)(\alpha^2+1)}{(n-\alpha^2)(\alpha^2+n-2)b^{(n-3)\gamma}}
\label{g}\end{equation}
As our solutions are neither asymptotically flat nor (A)dS, we get $g\geq 2$ in four dimension,
in contrast to asymptotically flat or (A)dS which have $g\leq 2$ in four dimensions \cite{Fro}.
In the absence of a non-trivial dilaton $(\alpha=\gamma=0)$, the gyromagnetic ratio reduces to:
\begin{equation}
g=n-1\label{gg}
\end{equation}

\section{Conclusions \label{IV}}
In this paper we construct the solutions of slowly
rotating black holes in $(n+1)$-dimensional
Brans-Dicke-Maxwell theory with a liouville-type potential in the
limit of a slow rotation parameter, with an arbitrary value of the coupling
constant $\omega$. Our solutions are neither asymptotically flat nor (A)dS,
 in contrast to the rotating black holes in the Einstein-Maxwell theory.
 The solutions are ill-defined for $\alpha=1$ and for $\alpha>1$
  we have cosmological horizons and there are no cosmological
 horizons for $\alpha<1$. In the latter case $(\alpha<1)$,
 we can have a black hole with inner and outer
 event horizons if $m>m_{ext}$, an extreme black hole
 if $m=m_{ext}$ and a naked singularity for $m<m_{ext}$.
 The cosmological horizons have a negative temperature for
 $\alpha>1$. We computed the Euclidean action and
  obtained the thermodynamics and conserved quantities. The
   temperature and entropy for this type of black hole were found to equal those in the static case to $O(a)$.
In addition  entropy does not follow the area law. Moreover we obtained
 angular momentum and gyromagnetic ratio for this rotating
 Brans-Dicke black hole. The gyromagnetic ratio is modified in this
 theory.


\end{document}